\documentclass[9pt,twocolumn,twoside]{pnas-new}

\templatetype{pnasresearcharticle} % Choose template 

\usepackage{amsmath}
\usepackage{amssymb}
\usepackage{graphicx}
\usepackage{float}
\usepackage{bm}
\usepackage{verbatim} 
\usepackage{mathrsfs}
\usepackage{xcolor}
\usepackage{dcolumn}
\usepackage{hyperref}
\usepackage[normalem]{ulem}

\renewcommand{\vec}[1]{\bm{#1}}
\newcommand{\Wi}{\mathrm{Wi}}
\renewcommand{\Re}{\mathrm{Re}}
\newcommand{\tr}{\mathrm{Tr} \,}

\newcommand{\alexander}[1]{\textcolor{black}{#1}}
\newcommand{\AM}[1]{\textcolor{black}{#1}}

\title{Purely elastic turbulence in pressure-driven channel flows}

\author[a]{Martin Lellep}
\author[b]{Moritz Linkmann} 
\author[a,1]{Alexander Morozov}

\affil[a]{School of Physics and Astronomy, The University of Edinburgh, James Clerk Maxwell Building, Peter Guthrie Tait Road, Edinburgh, EH9 3FD, United Kingdom}
\affil[b]{School of Mathematics and Maxwell Institute for Mathematical Sciences, The University of Edinburgh, Edinburgh, EH9 3FD, United Kingdom}

\leadauthor{Morozov} 

\significancestatement{Flows of complex fluids (such as polymers, colloids, emulsions, pastes, etc.) are abundant in everyday life. One can think of pouring shampoo or squeezing toothpaste from a tube, but also of fibre-spinning and plastic extrusion. Such flows often exhibit unexpected complexity with an archetypal example being elastic turbulence – chaotic motion exhibited by dilute polymer solutions. While its appearance is understood theoretically in flows with curved streamlines, little is known about elastic turbulence in straight pipes and channels. Here, we report the first simulations of elastic turbulence in a parallel flow geometry. Our results show that elastic turbulence is very different from its Newtonian counterpart and pave the way to understanding its mechanism.}

\authorcontributions{M. Linkmann and A.M. designed the research. All authors performed the numerical simulations and data analysis. M. Linkmann and A.M. wrote the paper.}
\authordeclaration{The authors declare no competing interests.}
\correspondingauthor{\textsuperscript{1}To whom correspondence should be addressed. E-mail: alexander.morozov@ed.ac.uk}

\keywords{Polymer solutions $|$ Viscoelastic flows  $|$ Elastic turbulence $|$ Direct numerical simulations} 

\begin{abstract}
Solutions of long, flexible polymer molecules are complex fluids that simultaneously exhibit fluid-like and solid-like behaviour. When subjected to an external flow, dilute polymer solutions exhibit elastic turbulence - a unique, chaotic flow state absent in Newtonian fluids, like water. Unlike its Newtonian counterpart, elastic turbulence is caused by polymer molecules stretching and aligning in the flow, and can occur at vanishing inertia. While experimental realisations of elastic turbulence are well-documented, there is currently no understanding of its mechanism. Here, we present large-scale direct numerical simulations of elastic turbulence in pressure-driven flows through straight channels. We demonstrate that the transition to elastic turbulence is sub-critical, 
giving rise to spot-like flow structures that, further away from the transition, eventually spread throughout the domain. We provide evidence that elastic turbulence is organised around unstable coherent states that are localised close to the channel midplane.
\end{abstract}

\dates{This manuscript was compiled on \today}
\doi{\url{www.pnas.org/cgi/doi/10.1073/pnas.XXXXXXXXXX}}

\begin{document}

\maketitle
\thispagestyle{firststyle}
\ifthenelse{\boolean{shortarticle}}{\ifthenelse{\boolean{singlecolumn}}{\abscontentformatted}{\abscontent}}{}

\dropcap{M}ost of the materials around us do not flow like Newtonian fluids do. Polymer rod-climbing \cite{Bird1987}, colloidal shear-thickening \cite{Wyart2014}, and shear-banding in worm-like micelles \cite{Lerouge2010} are just a few examples of such non-Newtonian behaviour that stems from the microstructure of these materials changing under external deformations. Despite their relevance to a wide spectrum of phenomena, ranging from microorganism propulsion \cite{Liu2021} and mechanics of biological tissues \cite{Chaudhuri2020} to polymer processing \cite{Denn2008}, time-dependent flows of non-Newtonian fluids are often beyond the reach of the current theory. An archetypal example of such unsteady flows is observed in dilute solutions of long, flexible polymer molecules \cite{Datta2022}. When subjected to a steady external forcing, polymer solutions exhibit instabilities to unsteady, chaotic flows often referred to as {\it elastic turbulence} \cite{Groisman2000}. Unlike its Newtonian counterpart, which is driven by fluid inertia, elastic turbulence originates in strongly anisotropic elastic stresses developing under external forcing, and can occur even in the absence of inertia \cite{Larson2000,Steinberg2021}. The transition to elastic turbulence is then controlled by a dimensionless parameter, the Weissenberg number $\Wi$ - a product of the applied velocity gradient scale and the polymer relaxation time, that determines the strength of the elastic stresses and plays the same role that the Reynolds number, $\Re$, does in Newtonian turbulence.

The route to elastic turbulence is best documented in model geometries with curved streamlines, where anisotropic elastic `hoop' stresses lead to linear instabilities \cite{McKinley1996} followed by a transition to chaotic flows at higher $\Wi$ \cite{Groisman2004,vanBuel2002}. Much less is known about flows of dilute polymer solutions in straight channels and pipes. Parallel shear flows of model polymer fluids are linearly stable under the experimentally relevant conditions \cite{Sanchez2022}, however, recent experiments present strong evidence of the presence of elastic turbulence in parallel shear geometries \cite{Bonn2011,Pan2013,jha2020preprint}. Indeed, it was proposed that the transition to elastic turbulence in such geometries is a `bifurcation from infinity' \cite{Rosenblat1979}, i.e. it requires a finite-amplitude flow perturbation \cite{Morozov2007}.   

Recently, the first steps were made in confirming this instability scenario theoretically. The development was spurred by the discovery of a linear instability in parallel shear geometries \cite{Khalid2021} that only exists in a remote part of the parameter space, with $\Wi\sim O(10^3)$ and at extremely low polymer concentrations. Although \alexander{this linear instability} might not be accessible experimentally and is not directly responsible for triggering elastic turbulence, it was shown to give rise to two-dimensional nonlinear travelling-wave solutions \cite{Page2020,Buza2021}. When continued outside this part of the parameter space, the travelling-wave solutions persist even in the absence of the underlying linear instability and can be found for a broad range of experimentally relevant parameters \cite{Morozov2022}. Below, we refer to these as `narwhal' solutions, reflecting the characteristic spatial profile of the associated polymer stretch. Although they resemble some features observed in experiments \cite{Bonn2011,Pan2013,jha2020preprint}, the `narwhal' states cannot be identified with elastic turbulence as they are not chaotic.

Despite these advances, it is presently not known how elastic turbulence is sustained. The main obstacle to further progress is the lack of understanding of the three-dimensional spatial distribution of the polymeric stresses and the associated fluid velocity of elastic turbulence, and how they evolve in time. In the absence of experimental measurements, it is natural to rely on direct numerical simulations to provide structural understanding of the flow, and this route has proved to be very fruitful in studying Newtonian turbulence \cite{Graham2021}. However, simulations of polymeric flows suffer from numerical instabilities and are notoriously difficult to perform \cite{Owens2002}. Despite a significant effort, there are no reports of successful simulations of three-dimensional elastic turbulence in confined geometries in the absence of a linear instability.

\begin{figure*}[!ht]
\includegraphics[scale=1]{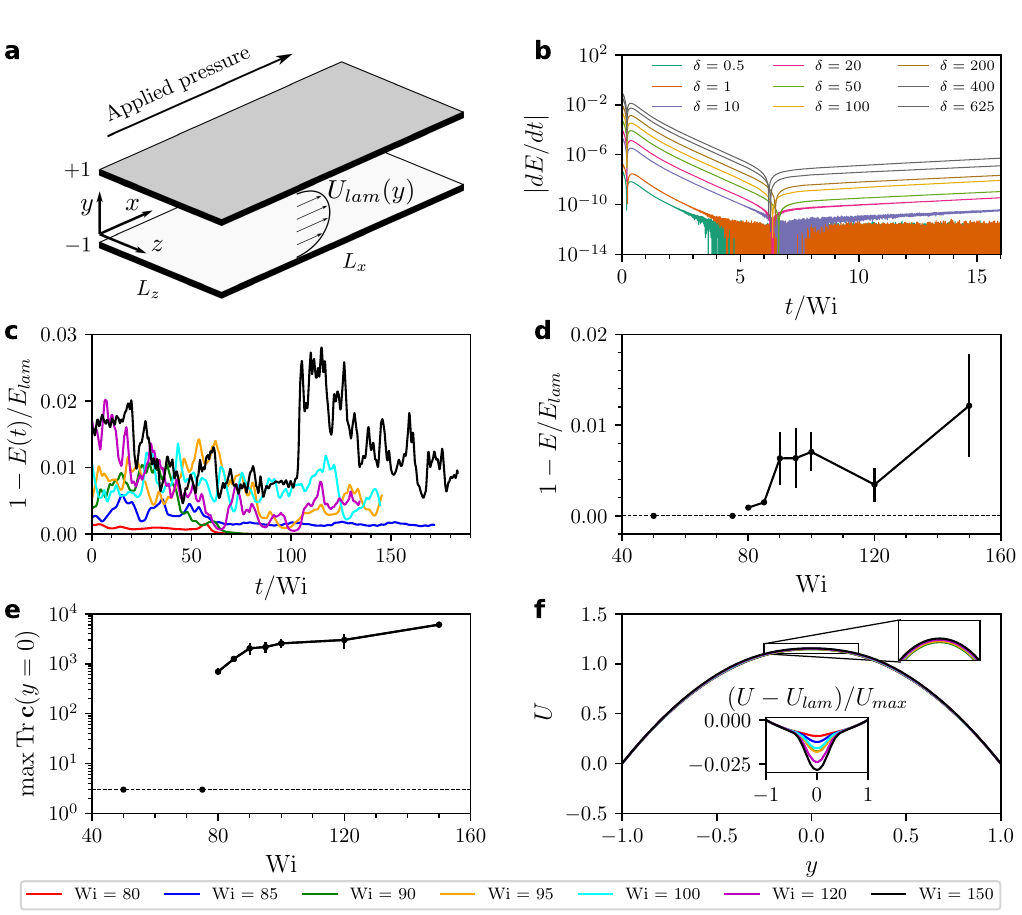}
\caption{{\bf Characterising the transition to elastic turbulence.} 
{\bf a}, Flow geometry. 
{\bf b}, Temporal evolution of the time derivative of the kinetic energy in simulations at $\Wi=150$ perturbed by random noise of amplitude $\delta$ (Materials and Methods). For small $\delta$, the derivative reaches zero up to machine precision, indicating relaminarisation, while at larger values of $\delta$ it grows in time. The subsequent time evolution of the run with the largest noise amplitude (SI Appendix Fig.S3a) shows that it develops into a fully chaotic state.
{\bf c}, Temporal evolution of the kinetic energy normalised by its laminar value. Time is measured in terms of polymer relaxation time $\lambda$ ($t/\Wi$ in our dimensionless units).
{\bf d} and {\bf e}, Bifurcation-from-infinity: the finite-amplitude jumps in the kinetic energy ({\bf d}) and the midplane polymer stretch ({\bf e}) from their laminar values.
{\bf f}, Mean streamwise velocity profiles. (Inset) Deviation of the mean streamwise velocity from its laminar profile.
}
\label{fig1}
\end{figure*}

To fill this gap, here we report direct numerical simulations of elastic turbulence in pressure-driven flow of a model dilute polymer solution. We study a model polymer fluid in a three-dimensional straight channel with $x$, $y$, and $z$ being Cartesian coordinates along the streamwise, wall-normal, and spanwise directions, respectively. The fluid fills the gap between two parallel plates and is driven by a constant pressure gradient applied externally \alexander{(Fig.\ref{fig1}a)}. We employ periodic boundary conditions in the $x$- and $z$-directions, with $L_x$ and $L_z$ denoting the simulation box lengths along the respective axes. A dilute polymer solution is modelled by the simplified Phan-Thien-Tanner (sPTT) constitutive relation \cite{PhanThien1977} given by 
\begin{align}
\label{eq:ptt} 
& \frac{\partial {\bm c}}{\partial t} + {\bm v}\cdot\nabla{\bm c} - \left(\nabla {\bm v}\right)^T\cdot{\bm c} - {\bm c}\cdot\left(\nabla {\bm v}\right) = \kappa \nabla^2 {\bm c} \nonumber \\
& \qquad\qquad\qquad\qquad\qquad\qquad - \frac{{\bm c}-\mathbb{I}}{\Wi}\Bigg[ 1-3\,\epsilon + \epsilon\, \mathrm{Tr} {\bm c}\Bigg],  \\
& \label{eq:ns} 
\frac{\partial {\bm v}}{\partial t} + {\bm v}\cdot\nabla{\bm v}  =
 -\nabla p + \frac{\beta}{\Re} \nabla^2{\bm v} + \frac{(1-\beta)}{\Re\,\Wi}\nabla\cdot{\bm c} + \frac{2}{\Re}\hat{\bf x}, \\
&\label{eq:incomp} 
\qquad\qquad\qquad\qquad  \nabla\cdot \vec{v} = 0,
\end{align}
where $\bm c$ is the polymer conformation tensor, $\bm v$ is the fluid velocity, $p$ is the pressure, and $\hat{\bf x}$ denotes a unit vector in the streamwise direction. 
Equations are rendered dimensionless by using $d$, $\mathcal{U}$, $d/\mathcal{U}$, and $\eta_p \mathcal{U}/d$, and $(\eta_s+\eta_p)\mathcal{U}/d$ as the units of length, velocity, time, stress, and pressure, respectively. Here, $d$ is the half distance between the channel plates, $\eta_s$ and $\eta_p$ are the solvent and polymeric contributions to the viscosity, and $\mathcal{U}$ is the maximal value of the laminar fluid velocity of a Newtonian fluid with the viscosity $\eta_s+\eta_p$ at the same value of the applied pressure gradient. The flow is characterised by the following dimensionless quantities: the Reynolds number, $\Re=\rho\,\mathcal{U} d/(\eta_s+\eta_p)$, the Weissenberg number, $\Wi=\lambda\,\mathcal{U} / d$, the viscosity ratio that acts as a proxy for the polymer concentration, $\beta = \eta_s/(\eta_s + \eta_p)$, the parameter controlling shear-thinning in the sPTT model, $\epsilon$, and the stress diffusivity, $\kappa$. Here, $\rho$ is the density of the fluid, and $\lambda$ is its Maxwell relaxation time. 
Throughout this work, we set $L_x=L_z=10$, $\Re=10^{-2}$, $\beta=0.8$, $\epsilon=10^{-3}$, and $\kappa=5\cdot 10^{-5}$. \AM{The value of $\kappa$ is estimated based on the diffusivity of a polymer molecule, $D\sim 1\mu$m$^2$/s, 
the relaxation time, $\lambda\sim 10$s, and the channel width, $d\sim 100\mu$m, as discussed in Ref. \cite{Morozov2022}.}
The fluid velocity obeys the no-slip boundary conditions, $\bm v(x,y=\pm 1,z,t) = 0$. The boundary conditions for the conformation tensor are obtained by requiring that ${\bm c}(x,y=\pm 1,z,t)$ is equal to the value obtained at the boundaries from \eqref{eq:ptt} with $\kappa=0$ \cite{Thomas2006}.

\alexander{For all $\Wi$ studied here, the laminar flow is linearly stable \cite{Sanchez2022}. To study its non-linear stability, we introduce a random perturbation of magnitude $\delta$ (Materials and Methods), and follow its time evolution. For sufficiently low Weissenberg numbers, $\Wi<80$, simulations return to the laminar state independent of the perturbation magnitude $\delta$. For $\Wi\ge80$, there exists a finite value of $\delta$, which is required to destabilise the flow (Figs.\ref{fig1}b and SI Appendix Fig.S3a). Following the non-linear evolution of the flow in this regime, we observe irregular temporal oscillations of the kinetic energy (Fig.\ref{fig1}c) commensurate with unsteady, chaotic dynamics observed in elastic turbulence \cite{Groisman2000}. These oscillations persist for many polymer relaxation times, measured in terms of $t/\Wi$ in our units (Figs.\ref{fig1}c), signifying the presence of non-linear, chaotic flow states for $\Wi\ge80$. The transition to purely elastic turbulence thus follows a bifurcation-from-infinity scenario \cite{Rosenblat1979,Morozov2007}, with finite-amplitude jumps in the observables separating the two states. \AM{The amplitude of the kinetic energy jump, however, is small (Fig.\ref{fig1}d), in line with a weakly sub-critical bifurcation that has recently been reported in experiments \cite{Pan2013,Qin2017}. We note that it can easily be mistaken for a linear instability.} The amplitude of the jump of the midplane polymer stretch, on the other hand, is significantly higher (Fig.\ref{fig1}e), consistent with the polymer stress being the key dynamical variable of purely elastic turbulence.}

\alexander{Notably, close to the onset of elastic turbulence, the long time evolution of the non-linear flow states is followed by a sudden return to the laminar state (see Supplementary Movie 1), and we observed such events for $\Wi=80$ and $\Wi=90$, but not for $\Wi=85$ (Fig.\ref{fig1}c). Sudden relaminarisation events are a key characteristics of the transition scenario in linearly stable Newtonian parallel shear flows \cite{Barkley2016}, indicative of a fractal laminar-turbulent boundary \cite{Moehlis2004}, finite turbulent lifetimes \cite{Hof2006}, and localised flow structures \cite{Avila2011}. Our observations, supported by further evidence below and by recent experimental reports of sudden splitting of localised structures in elastic pressure-driven channel flows \cite{Shnapp2022b}, suggest a similar transition scenario for linearly stable purely elastic flows.
}

\begin{figure*}[!ht]
\includegraphics[scale=1]{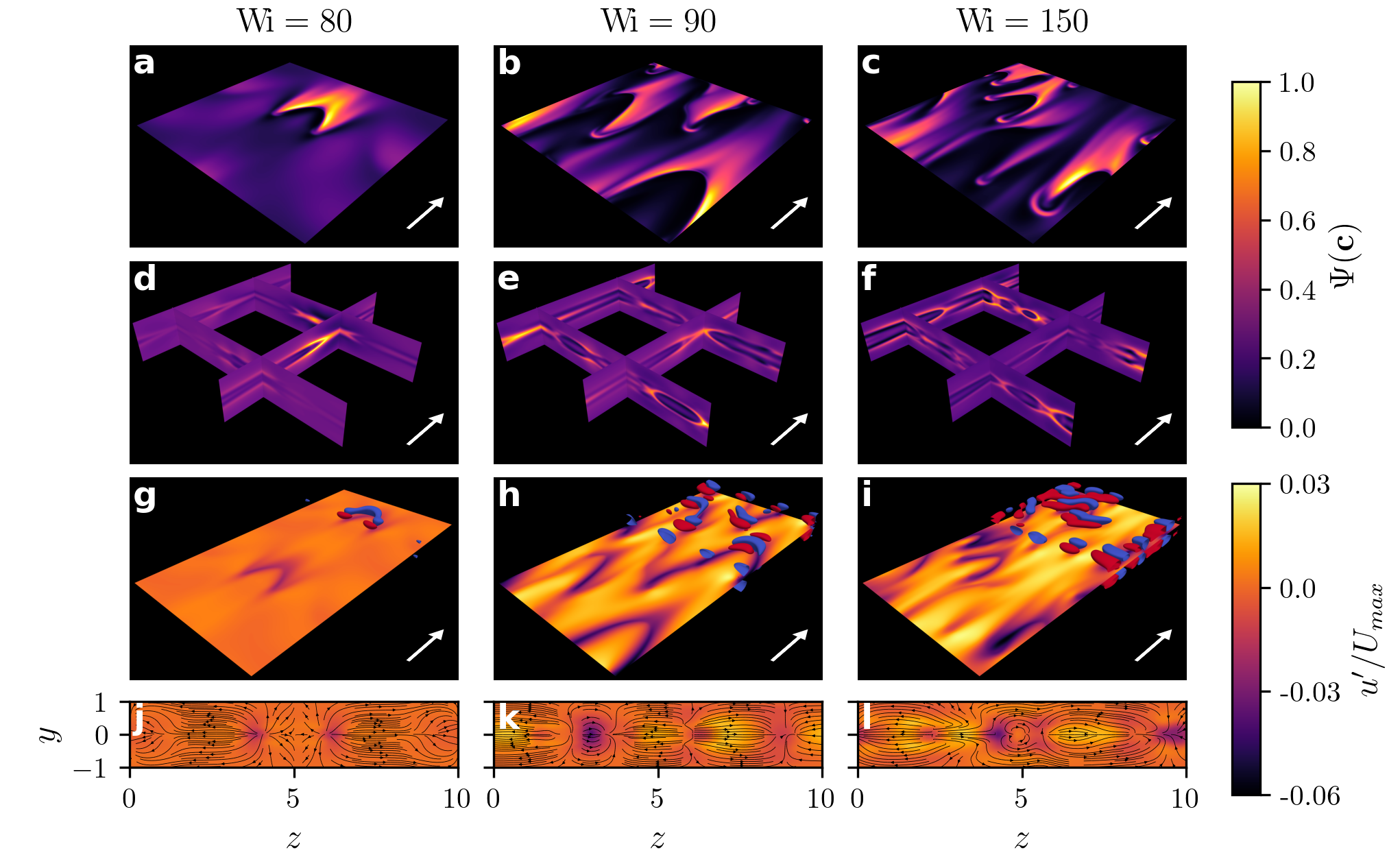}
\caption{{\bf Instantaneous spatial structure of elastic turbulence at various $\Wi$.} The mean flow direction is indicated by the white arrows.
{\bf a-c}, Midplane profile of the relative polymer stretch, $\Psi({\bm c})$ (Materials and Methods). {\bf d-f}, Vertical slices of $\Psi({\bm c})$ at several positions in the channel demonstrating variations in the $y$-direction. {\bf g-i}, The midplane profile of $u'/U_{max}$ and the isosurfaces of the Q-criterion (Materials and Methods) visualised for values $- 0.002$ and $0.002$ in blue and red, respectively. The simulation domain is repeated twice in the streamwise direction to improve readability of the figure. {\bf j-l}, Streamwise slices of $u'/U_{max}$ with the in-plane streamlines at $x=5$.
}
\label{fig2}
\end{figure*}

Our simulations provide direct access to the spatial distribution of the flow velocity and the polymeric stress and allow us to understand the structural features of elastic turbulence in straight channels. For all $\Wi$ studied, we observe that the \alexander{largest deviations of the polymer stress from its laminar profile} were mainly localised in a thin sheet around the channel centreline (Fig.\ref{fig2}a-c), while the flow is almost laminar close to the walls. This is in a stark contrast with Newtonian turbulence that exhibits the strongest fluctuations close to confining walls \cite{Smits2011}. 
For low values of the Weissenberg number, close to the saddle-node bifurcation at $\Wi \approx 80$, the polymer extension presents a spatially localised profile, similar to turbulent puffs and spots observed in Newtonian pipe and channel flows \cite{Tuckerman2020}, respectively. For larger values of $\Wi$, these localised structures proliferate throughout the domain in what resembles a percolation transition \cite{Barkley2016}, exhibiting chaotic splitting and merging (see Supplementary Movies 1-3). \alexander{Such events are reflected in the strongly intermittent temporal behaviour of the kinetic energy (Fig.\ref{fig1}c) and the midplane polymer stretch (SI Appendix Fig.S3a).}
Throughout these dynamics, the stress profile in the streamwise-wall-normal plane (Fig.\ref{fig2}d-f) preserves the overall features of the two-dimensional `narwhal' states. This suggests that although they are linearly unstable \cite{Lellep2023}, the subcritical `narwhal' states organise the three-dimensional chaotic dynamics close to the onset of elastic turbulence. 

\begin{figure*}[!ht]
\includegraphics[scale=1]{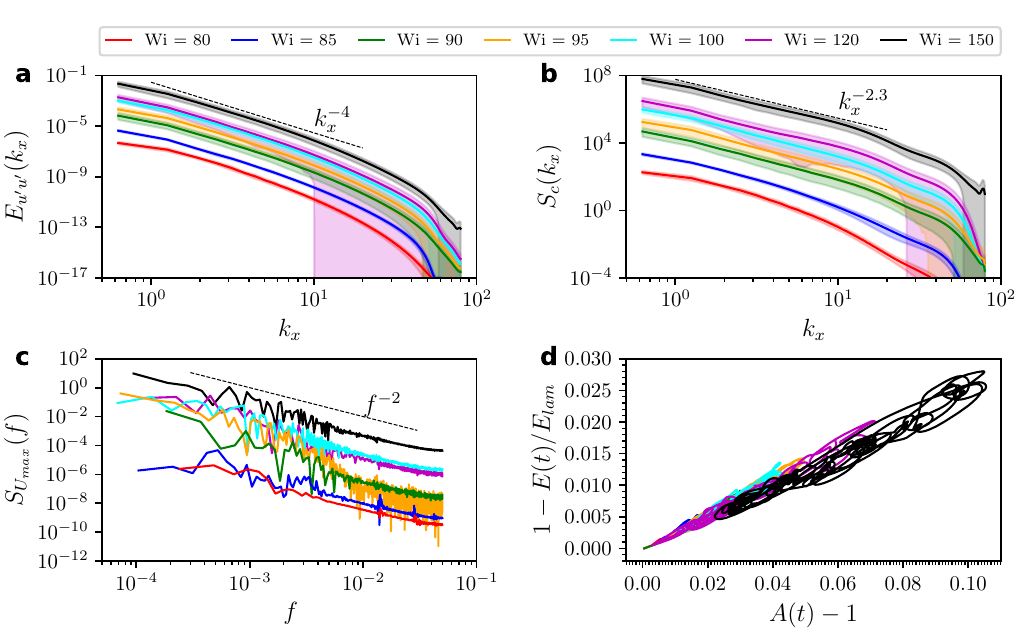}
\caption{ {\bf Spatio-temporal properties of elastic turbulence at various $\Wi$.} 
One-dimensional spatial spectra of the streamwise velocity component ({\bf a}) and the trace of the conformation tensor ({\bf b}). {\bf c}, Temporal spectrum of the centerline velocity. In ({\bf a-c}), the dashed lines indicate tentative power-law scaling laws, the shaded regions indicate one standard deviation, and the data have been shifted vertically by the same offset to improve the readability of the figure. {\bf d}, Phase-space projections of the chaotic trajectories. }
\label{fig3}
\end{figure*}

We observe that the most prominent velocity features of elastic turbulence are associated with the streamwise velocity component. Since the chaotic stress dynamics are largely confined to a region around the centreline, the deviations of the mean streamwise velocity $U(y)$ from its laminar profile are only noticeable in the middle of the channel (Fig.\ref{fig1}f) and constitute a few percent of the laminar centreline value. These observations are consistent with previous experimental measurements \cite{Pan2013} that identified the centreline velocity fluctuations as an observable sensitive to the onset of elastic turbulence in channel flows. The spatial distribution of the streamwise velocity on the midplane (Fig.\ref{fig2}g-i) exhibits low- and high-velocity streaks similar to the structures reported in recent experiments \cite{jha2020preprint}. In contrast to Newtonian turbulence, the presence of streaks is not associated with the streamwise-oriented vortices, characteristic of the near-wall cycle \cite{Waleffe1997} (Fig.\ref{fig2}j-l). Instead, we observe that the midplane streaky features and the chevron-like stress profile are reminiscent of the flow patters in viscoelastic Kolmogorov flow \cite{Garg2021}, and we hypothesise that a shear-layer instability plays a role in sustaining three-dimensional elastic turbulence in channel flows. The final velocity features of elastic turbulence in our simulations are the spanwise-oriented vortices placed symmetrically above and below the midplane (Fig.\ref{fig2}g-i). Such structures are remarkably similar to the vortices reported in elasto-inertial turbulence at high $\Re$ \cite{Samanta2013} and, thus, our simulations provide support to recent claims that elastic and elasto-inertial turbulence have the same physical origins \cite{Sanchez2022,Dubief2023} and that there is a continuous connection between these phenomena in the $\Re-\Wi$ space  \cite{Khalid2021}.

To analyse the statistical features of the ET dynamics, we study one-dimensional spatial spectra of the streamwise velocity fluctuations (Fig.\ref{fig3}a) at the channel midplane. We observe that the kinetic energy spectra are broadly consistent with the $E_{u'u'}\propto k_x^{-4}$ behaviour, with the extent of the scaling region increasing with $\Wi$ (SI Appendix Fig.S1).  This power law, considered to be a hallmark of polymer-induced chaotic flows \cite{Fouxon2003,Dubief2013,Garg2021,Steinberg2021}, indicates that the velocity field of elastic turbulence observed in our simulations is spatially smooth, unlike its Newtonian counterpart \cite{Onsager1949}.  As can be further seen from Fig.\ref{fig3}a, the error bars, indicative of one standard deviation, become large at high wavenumbers $k_x$. We observe that there are instances in time where velocity and polymer stress fluctuations are confined to large and intermediate scales, with the corresponding spectra dropping off steeply at sufficiently large $k_x$ (SI Appendix Fig.S2). At other instances in time, the small scales fluctuate intensely.  This indicates an intermittent breakdown of small-scale dynamics. We hypothesise that the dynamics at small scales are important to sustaining elastic turbulence; this, in turn, would explain why our simulations require large spatial resolution to successfully capture polymer-induced chaotic flows.
Temporal statistical features are captured by the spectra of the centerline velocity (Fig.\ref{fig3}c), that scale with the frequency $f$ as $S_{U_{\rm max}}(f) \propto f^{-2}$.  \AM{Spectral exponents in line with this value have been measured experimentally around the transition to turbulence in viscoelastic channel flow \cite{Pan2013,Qin2017,Shnapp2022}.}

At small $\Re$, the main dynamical variable of elastic turbulence is the polymer conformation tensor, while the velocity field is linearly enslaved to its time-evolution by virtue of Stokes' equation. To study its statistical behaviour, we plot one-dimensional spectra of $\tr\bm{c}$ (Fig.\ref{fig3}b) and observe that $S_c\propto k_x^{-2.3}$. A spectral exponent of $-2$ is readily obtained by dimensional analysis from the velocity spectrum reported above and Stokes' equation. However, a small but finite deviation of the spectral exponent of the stress fluctuation from predictions based on dimensional analysis indicates the presence of additional effects determining the statistics of the stress fluctuations. This motivates further analyses, for instance concerning statistical self-similarity as a function of Wi.

Our simulations also provide indirect evidence that the chaotic dynamics of ET close to its onset is organised by unstable coherent structures, such as relative periodic orbits. In Fig.\ref{fig3}d, we plot the phase-space projections of high-dimensional chaotic trajectories corresponding to the time-series of the kinetic energy shown in Fig.\ref{fig1}c onto a two-dimensional subspace spanned by the deviation of the kinetic energy from the laminar profile, $1-E(t)/E_{lam}$, and the normalised mean of the trace of the polymer stress, $A(t)$ (Materials and Methods). We observe that for each $\Wi$ the phase-space trajectories evolve around points and closed curves in the two-dimensional subspace for long times. Similar observations in Newtonian flows were shown to indicate the presence of coherent states in the vicinity of the trajectories \cite{Kawahara2001,Hof2006, Avila2011,Linkmann2015}. As noted above, the spatial stress configuration resembling the two-dimensional `narwhal' states \cite{Morozov2022} is clearly visible in Fig.\ref{fig2}d-f, suggesting that, although being linearly unstable, these states form a key part of the ET coherent structures. Finally, we note that our simulations exhibit a high degree of spatio-temporal intermittency, with the number of spatially localised states fluctuating in time (Supplementary Movies 2 \& 3). These dynamics are a hallmark of the dynamical process based on splitting and decaying of spatially localised structures. Future work is required to demonstrate whether this process belongs to the directed percolation universality class \cite{Hinrichsen2000}, as is the case in Newtonian parallel shear flows \cite{Lemoult2016,Sano2016,Barkley2016}. 

%\newpage

\matmethods{

\subsection*{Definition of the observables}

The velocity field and the conformation tensor are decomposed into the mean profile and fluctuations,
\begin{align}
& \bm{v}  = \langle \bm{v}(y) \rangle_{x,z,t} + \bm{v}'(x,y,z,t), \\
& \bm{c}  = \langle \bm{c}(y) \rangle_{x,z,t} + \bm{c}'(x,y,z,t), 
\end{align}
where $\langle \dots \rangle_{x,z,t}$ denotes the spatial average along the streamwise and spanwise directions, and the temporal average in the statistically stationary state. In particular, we study the mean streamwise velocity profile, $U(y)$, and the instantaneous deviation of the streamwise velocity from its mean profile, $u'$. The maximal value of the mean streamwise velocity is denoted by $U_{max}$ and always corresponds to $U(y=0)$. The laminar values of the streamwise velocity and the conformation tensor are denoted by $U_{lam}(y)$ and $\bm{c}_{lam}(y)$, respectively. 

The instantaneous ratio of the kinetic energy to its laminar value is defined through
\begin{align}
\frac{E(t)}{E_{lam}} = \frac{1}{L_x L_z}\frac{\int_0^{L_x} \int_{-1}^1 \int_0^{L_z} dx \ dy \ dz \ \bm{v}(x,y,z,t)^2}{\int_{-1}^1 dy  \ U_{lam}(y)^2}.
\end{align}
This quantity is used in Fig.\ref{fig1}b (without the normalisation), Fig.\ref{fig1}c, and Fig.\ref{fig3}d, while Fig.\ref{fig1}d shows its value averaged over time in the statistically stationary state. Similarly, the instantaneous ratio of the volume-averaged polymer stretch to its laminar value is defined through
\begin{align}
A(t) = \frac{1}{L_x L_z}  \frac{\int_0^{L_x} \int_{-1}^1 \int_0^{L_z} dx \ dy \ dz \ \tr \bm{c} (x,y,z,t)}{\int_{-1}^1 dy \ \tr \bm{c}_{lam}(y)}.
\end{align}

To aid the visualisation of the flow structures, in Fig.\ref{fig2} we introduce the relative polymer stretch, defined as
\begin{align}
\Psi(\mathbf{c}) = \frac{ \tr\!\left(\bm{c}- \bm{c}_{lam}\right) - \rm{min}\, \tr\!\left(\bm{c}- \bm{c}_{lam}\right)  } { \rm{max}\,  \tr\!\left(\bm{c}- \bm{c}_{lam}\right) - \rm{min}\,  \tr\!\left(\bm{c}- \bm{c}_{lam}\right) },
\end{align}
that allows for comparison between simulations at different $\Wi$.

%The increase in the friction associated with elastic turbulence is characterised by the ratio of the Darcy friction factor to its laminar value, defined as
%\begin{align}
%%\frac{f}{f_{lam}} = \frac{U_{b, lam}^2}{U_b^2},
%\frac{U_{b, lam}^2}{U_b^2},
%\end{align}
%where $U_b = \frac{1}{2}\int_{-1}^{1} dy \ U(y)$ and $U_{b, lam} = \frac{1}{2}\int_{-1}^{1} dy \ U_{lam}(y)$ are the bulk velocity and its laminar value, respectively.
%
One-dimensional spatial spectra of the streamwise velocity and the trace of the conformation tensor fluctuations at the channel centre-plane are defined as 
\begin{align}
& E_{u'u'}(k_x) = \left \langle \int dk_z \left |\hat{u}'(k_x,y=0,k_z)\right |^2  \right \rangle_t, \\
& S_{c}(k_x) = \left \langle \int dk_z \left |{\tr \hat{\bm{c}}}'(k_x,y=0,k_z)\right |^2 \right \rangle_t, 
\end{align}
where $\langle\dots\rangle_t$ denotes the temporal average calculated in the statistically stationary state, and $\hat{\cdot}$ denotes the two-dimensional Fourier transform along the $x$- and $z$-directions with $k_x$ and $k_z$ being the corresponding components of the wavevector. The temporal spectra of the centerline velocity $U_{max}$ are calculated as
\begin{equation}
    S_{U_{max}}(f) = \left |\mathcal{F}[U(y=0)](f)\right |^2  \ , 
\end{equation}
where $f$ is the frequency, and $\mathcal{F}$ denotes the temporal Fourier transform.

\subsection*{Q-criterion}

The Q-criterion \cite{Hunt1988} is a Galilean-invariant vortex identification measure that is based on the second invariant of the velocity gradient tensor
\begin{align}
Q = \frac{1}{2} \left ( \Omega_{ij}\Omega_{ji} - S_{ij}S_{ij} \right), 
\end{align}
where $S_{ij} = (\partial_j v_i + \partial_i v_j)/2$ and $\Omega_{ij} = (\partial_j v_i - \partial_i v_j)/2$ are the strain rate and the vorticity tensor, respectively, and indices denote Cartesian components of the vectors. A vortex is defined as a region in the flow where $Q > 0$.

\subsection*{Direct numerical simulations}

Equations \eqref{eq:ptt}-\eqref{eq:incomp} are solved numerically with an in-house MPI-parallel fully-dealiased pseudo-spectral code developed within the Dedalus framework \cite{Burns2020}. The velocity, conformation tensor, and pressure are represented by a Fourier-Chebyshev-Fourier spectral decomposition along the streamwise, wall-normal, and spanwise directions, respectively, with the spectral resolution being given by the number of modes in each direction $(N_x,N_y,N_z)$. Production runs are carried out using $N_x=N_z=256$ and $N_y=1024$, comprising approximately $7\cdot 10^8$ degrees of freedom. Temporal discretisation employs the semi-implicit backward differentiation scheme of order four \cite{Wang2008} with time step of 0.005. All simulations have been evolved for at least $130$ polymer relaxation times ($10^4$-$10^5$ time units depending on $\Wi$) in a statistically stationary state. Data are sampled at intervals of $10$ time units for full-state snapshots and $0.1$ time units for the kinetic energy.  A summary of all simulations is provided in SI Appendix Table S1. A typical production run was carried out on 16384 cores for approximately 260 hours.

We employed three types of initial conditions in our simulations. In Fig.\ref{fig1}b, we perturbed the laminar state by adding $\delta \vert \xi \vert$ to the $c_{xx}$ component of the conformation tensor. Here, $\xi$ is a position-dependent random Gaussian noise with zero mean and unit variance, and $\delta$ is the amplitude of the noise. The absolute value ensured that the perturbation preserved the positive-definiteness of the conformation tensor. This strategy resulted in very long runs to reach the statistically stationary state, we have only used it in Fig.\ref{fig1}b to demonstrate the existence of a finite-amplitude threshold, and in \AM{SI Appendix Fig.S3}. For the second strategy, we embedded the two-dimensional `narwhal' states \cite{Morozov2022} into our three-dimensional domain with a small amount of random noise added to break the translational symmetry along the spanwise direction. This strategy was employed for $\Wi=100$, and the quick destabilisation of the translationally invariant `narwhal' state is visible at early times in \AM{Supplementary Movie 2}. All other simulations presented here were started from the chaotic states obtained from the previous runs at a new value of $\Wi$. 

To demonstrate that different initial conditions produce the same statistical steady states, we continued the simulation at $\Wi=150$ with the highest noise amplitude ($\delta=625$ in Fig.\ref{fig1}b) until it reached a statistically steady-state. Additionally, to demonstrate convergence with respect to spatial discretisation, we repeated the same run at a reduced spatial resolution $N_x=N_z=128$ and $N_y=512$. In SI Appendix Fig.S3a, these runs are compared to the $\Wi=150$ simulation from Fig.\ref{fig1}c. Although the dynamics are strongly intermittent in all cases, visual inspection confirms convergence to a statistically steady state. This observation is further supported by SI Appendix Fig.S3b where we find no statistically significant differences between the midplane polymer stretch in these simulations.  
}

\showmatmethods{} % Display the Materials and Methods section

\acknow{We thank Keaton Burns and Davide Marenduzzo for valuable suggestions. This work used the ARCHER2 UK National Supercomputing Service (https://www.archer2.ac.uk). Support from the UK Turbulence Consortium (EPSRC grants EP/R029326/1 and EP/X035484/1) and ARCHER2 team are gratefully acknowledged.
Martin Lellep is supported by the German Academic Scholarship Foundation (Studienstiftung des deutschen Volkes). We acknowledge financial support from the Priority Programme SPP 1881 ``Turbulent Superstructures" of the Deutsche Forschungsgemeinschaft (DFG) under grant Li3694/1. For the purpose of open access, the authors have applied a Creative Commons Attribution (CC BY) licence to any Author Accepted Manuscript version arising from this submission.
}
\showacknow{} 

% Bibliography
\bibliography{nature}

\end{document}

%% Comment out or remove this line before generating final copy for submission; this will also remove the warning re: "Consecutive odd pages found".
%\instructionspage  

\maketitle

%% Adds the main heading for the SI text. Comment out this line if you do not have any supporting information text.
%\SItext

%\begin{figure*}[ht]
%\begin{center}
%\includegraphics[scale=1]{frictionfactor.pdf}
%\caption{The ratio of the Darcy friction factor to its laminar value (Materials and Methods) as a function of $\Wi$.}
%\end{center}
%\end{figure*}

\begin{figure*}[ht]
\includegraphics[scale=1]{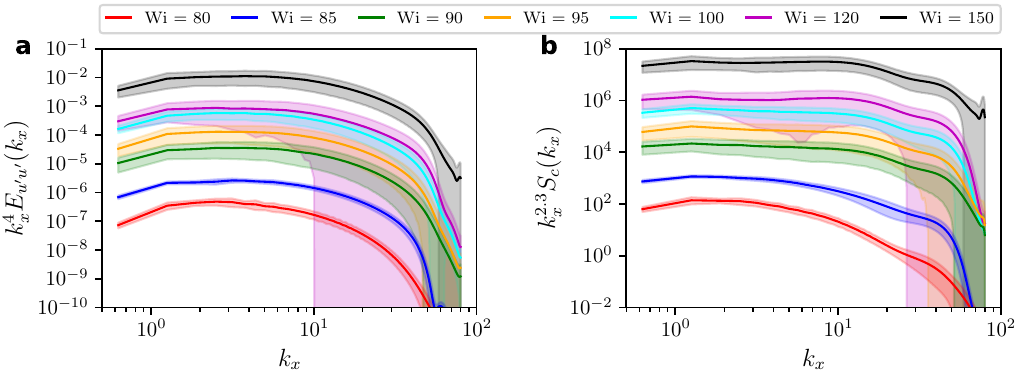}
\caption{{\bf Compensated spectra at different Weissenberg numbers.} {\bf a}, One-dimensional spectrum of the streamwise velocity component. {\bf b}, One-dimensional spectrum of the trace of the conformation tensor. The shaded regions indicate one standard deviation. Data have been shifted vertically by the same offset in both subfigures to improve the readability of the figure.}
\end{figure*}

\begin{figure*}[ht]
\begin{center}
\includegraphics[scale=1]{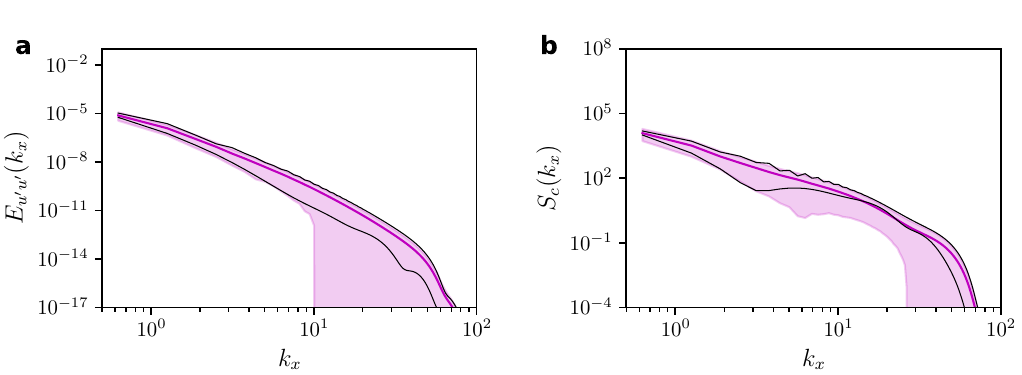}
\caption{ {\bf Instantaneous spectra for $\Wi = 120$.}  {\bf a}, One-dimensional spectrum of the streamwise velocity component.  {\bf b}, One-dimensional spectrum of the trace of the conformation tensor. The shaded regions indicate one standard deviation. The violet lines are ensemble-averaged spectra, while the black lines are representative instantaneous realisations.}
\end{center}
\end{figure*}

\begin{figure*}[ht]
\begin{center}
\includegraphics[scale=1]{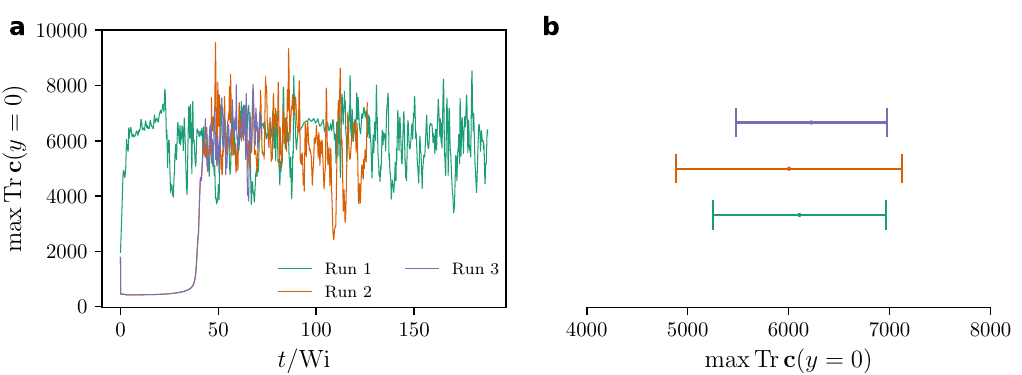 }
\caption{ {\bf Studying the influence of the initial condition and spatial resolution for $\Wi = 150$.}  {\bf a}, Time evolution of the midplane polymer stretch. Run 1 employed a configuration taken from the $\Wi=100$ simulation as its initial condition. The data is re-plotted from Fig.1c of the main text. Runs 2 and 3 were started from the laminar state perturbed by random noise with the amplitude $\delta=625$ (Materials and Methods). The small-$t$ data of Run 2 are plotted in Fig.1b. Runs 1 and 2 employed the production spatial resolution, while Run 3 employed a reduced (halfed) resolution (Materials and Methods).  {\bf b}, Statistical analysis of the data from (a) evaluated in the statistical steady state. The circles denote the averages, while the lines indicate one standard deviation. In the absence of statistically-significant or systematic discrepancies between the runs, we conclude that all three simulations have converged to the same chaotic steady state.}
\end{center}
\end{figure*}

%%% Add this line AFTER all your figures and tables
\FloatBarrier

\begin{table*}[ht]
    \centering
     \caption{{\bf Parameters and key observables used in the statistical analyses.}
$E$ and $\langle \tr \bm{c}\rangle$ are the volume- and time-averaged kinetic energy and the trace of the conformation tensor, respectively, while
$E_{lam}$ and $\tr \bm{c}_{lam}$ are the corresponding laminar values. 
$t_{start}$ ($t_{stop}$) is the starting (stopping) time of the interval for the calculation of statistical observables and $M$ is the number of data samples used in the calculation of statistical observables.}
 \begin{tabular}{cccccccc}
                $\Wi$ & $E$ & $E_{lam}$ & $\langle \tr \bm{c}\rangle$ & $\tr \bm{c}_{lam}$  & $t_{start} / Wi$ & $t_{stop} / Wi$ & $M$ \\
                \hline
                50  & 1.308 & 1.308  & 1213  & 1213 &  - & - &  - \\
                75  & 1.357  & 1.357 & 1769  & 1769 &  - & -  &  - \\
                80  & 1.362  & 1.364 & 1882  & 1874 & 0 & 52 & 415 \\
                85  & 1.368  & 1.370 & 1990  & 1977 & 60 & 172 & 955 \\
                90  & 1.367  & 1.376 & 2127  & 2079 & 0 & 60 & 540 \\
                95  & 1.373  & 1.382 & 2231  & 2179 & 0 & 145 & 1374 \\
                100  & 1.377  & 1.387 & 2339  & 2277 & 0 & 145 & 1459 \\
                120  & 1.396  & 1.404 & 2688  & 2656 & 60 & 134 & 896 \\
                150  & 1.406  & 1.423 & 3353  & 3191 & 22 & 151 & 1935 \\
                \hline
        \end{tabular} 
%    \begin{tabular}{cccccccc}
%    $\Wi$ & $E$ & $E_{lam}$ & $\langle \tr \bm{c}\rangle$ & $\tr \bm{c}_{lam}$  & $t_{start} / Wi$ & $t_{stop} / Wi$ & $M$ \\
%     \hline
%    50  & 1.308 & 1.308  & 1213  & 1213 &  - & - &  - \\
%    75  & 1.357  & 1.357 & 1769  & 1769 &  - & -  &  - \\
%    80  & 1.352  & 1.364 & 1868  & 1874 & 0 & 52 & 415 \\
%    85  & 1.357  & 1.370 & 1974  & 1977 & 60 & 176 & 955 \\
%    90  & 1.357  & 1.376 & 2110  & 2079 & 0 & 60 & 540 \\
%    95  & 1.362  & 1.382 & 2213  & 2179 & 0 & 158 & 1374 \\
%    100  & 1.366  & 1.387 & 2320  & 2277 & 0 & 175 & 1459 \\
%    120  & 1.385  & 1.404 & 2667  & 2656 & 60 & 143 & 896 \\
%    150  & 1.396  & 1.423 & 3317  & 3191 & 5 & 75 & 1057 \\
%    \hline
%    \end{tabular}
   \end{table*}

\movie{Elastic turbulence at $\Wi=80$. The simulation rapidly converges to a localised turbulence structure that persists for a long time before suddenly relaminarising.}

\movie{Elastic turbulence at $\Wi=100$. The simulation is started from a two-dimensional `narwhal' state, translationally-invariant along the spanwise direction, perturbed by a small amount of noise. Early time evolution shows that this state is unstable and the simulation quickly reaches a chaotic steady state.}

\movie{Elastic turbulence at $\Wi=150$. A strongly intermittent simulation exhibiting splitting and merging of localised coherent structures.}